\documentclass{aa}
\usepackage{graphicx}
\usepackage{txfonts}

\usepackage{hyperref}
\usepackage[utf8]{inputenc}
\usepackage[T1]{fontenc}
\usepackage{bbold} 
\usepackage{siunitx} 

\begin{document}

   \title{Assessing the reliability of the Bisous filament finder}

   \author{{Moorits Mihkel} Muru\inst{\ref{TO}}
            \and
            Elmo Tempel\inst{\ref{TO}}
            }

   \institute{Tartu Observatory, University of Tartu, Observatooriumi 1, 61602 Tõravere, Estonia}\label{TO}

   \date{\today}

 
\abstract
{Recent years have given rise to numerous methods of detecting the cosmic web elements in the large-scale structure of the Universe. All of these methods describe more or less the same features, but each with its nuance. The Bisous filament finder is a stochastic tool for identifying the spines of filaments using galaxy positions.}
{This work provides an analysis of how the galaxy number density of the input data affects the filaments detected with the Bisous model and gives estimates of the reliability of the method itself to assess the robustness of the results.}
{We applied the Bisous filament finder to \textsc{MultiDark-Galaxies} data, using various magnitude cuts from the catalogue to study the effects of different galaxy number densities on the results and different parameters of the model. We compared the structures by the fraction of galaxies in filaments and the volume filled by filaments, and we analysed the similarities between the results from different cuts based on the overlap between detected filamentary structures. The filament finder was also applied to the exact same data 200 times with the same parameters to study the stochasticity of the results and the correlation between different runs was calculated.}
{Multiple samples show that galaxies in filaments have preferentially higher luminosity. We found that when a galaxy is in a filament there is a 97\% chance that the same galaxy would be in a filament with even more complete input data and about 85\% of filaments are persistent when detecting the filamentary network with higher-density input data. Lower galaxy number density inputs mean the Bisous model finds fewer filaments, but the filaments found are persistent even if we use more complete input data for the detection. We calculated the correlation coefficient between 200 Bisous runs on the exact same input, which is \num{0.98}.}
{This study confirms that increased number density of galaxies is important to obtain a more complete picture of the cosmic web. To overcome the limitation of the spectroscopic surveys, we will develop the Bisous model further to apply this tool to combined spectroscopic and narrow-band photometric redshift surveys, such as the J-PAS.}


\maketitle


\section{Introduction}

The concordance cosmology has been shown to predict the evolution of dynamically bound large-scale structure \citep{Shandarin1989}. Although simulations \citep[e.g.][]{Springel05} have confirmed the emergence of cosmic structures, similar to those observed, some cosmological parameters are still in need of fine-tuning, for example the so-called Hubble tension \citep{Mortsell2018, Feeney2019}. The parameters in the cosmological model affect the evolution of large-scale structure. Therefore being able to characterise the observable large-scale structure of the Universe helps narrow down the precise values of these cosmological parameters.

Further, numerous studies have shown how the local evolution of galaxies, groups, and clusters depends on the large-scale structure. For example, the environment affects the luminosity function \citep{Tempel2009, Tempel2011}, the evolution of galaxies \citep{Lietzen2012, Einasto2018}, the morphology \citep{Kuutma2017}, and the properties of central galaxies in groups \citep{Poudel2017}. Mapping the large-scale structure could prove to be useful for solving even larger problems, for example, the missing baryon problem \citep{Shull2012}, for which the large-scale structure could be used as a tracer to detect diffuse and hot gas in the intergalactic medium that is otherwise hard to detect \citep{Nevalainen15}. This only adds to the significance of being able to measure and quantify the large-scale structure of the Universe. 

Although some of the first studies of the large-scale structure were conducted decades ago \citep{Fall1980, Frenk1988}, the last several years have been fruitful for the topic in terms of the emergence of new approaches to detect the large-scale structure. There is no consensus on how to define the elements of the large-scale structure and that has given rise to a plethora of methods for finding these elements. \citet{Libeskind18} gives a short overview and compares some of the methods. These approaches are mainly divided into five classes: graph and percolation techniques \citep{Alpaslan2014}, stochastic methods \citep{Tempel14, Tempel16, Gonzalez2017}, geometric and Hessian-based methods \citep{ForeroRomero2009, Hoffman2012, Kitaura2012, Cautun2013, AragonCalvo2007, AragonCalvo2014}, topological methods \citep{AragonCalvo2010, Sousbie2011}, and phase-space methods \citep{Falck2012, Falck2015, Ramachandra2015}. In this work, we use the Bisous filament finder, which is specifically developed to find and characterise filamentary structures applied to observational data. The Bisous filament finder is a stochastic model, which is based on the marked point processes \citep{baddeley2015spatial} and aims to model the filamentary network underlying the distribution of galaxies. The mathematical basis of the method has been described and proven in \citet{Stoica2005a, Stoica2005b, Stoica2007a, Stoica2007b, Stoica2010, Stoica2014, Tempel14, Tempel16}.

As expected, methods constructed from entirely different concepts and algorithms return different results even from the same input data. But the results are similar enough, showing connected web-like structures, to be comparable. Metrics need to be devised to compare the results obtained from different methods. For example, to compare the detected filamentary network, \citet{Rost20} uses filament lengths distribution, galaxy overdensities in filaments, filament luminosities, mean galaxy luminosity in filaments, and luminosity function of galaxies in filaments. \citet{Libeskind18} compares various filament finding techniques using mass and volume filling fractions, densities of different structure elements, and mass functions. The former comparison uses observational data from Sloan Digital Sky Survey data release 12 (SDSS DR12) \citep{SDSS_III, SDSS_DR12} and the latter comparison dark matter-only simulation; both of these include the Bisous model in their comparison.

In addition to comparing with different methods, a method can be compared with itself using different configurations or method parameters. This paper aims to give an overview of how different parameters affect the filament extraction using the Bisous model. For the results with different parameters to be comparable with other methods, we use similar metrics as in \citet{Libeskind18}.

In contrast to deterministic methods, stochastic methods may get different results from identical input and configurations. But every meaningful result should converge over a number of iterations, which begs the question of how many iterations are sufficient and how the uncertainties scale with the number of iterations. This robustness is important for interpreting the results for a larger context or for deriving conclusions from these findings. This work studies the robustness of the Bisous filament finder to assess the possible uncertainties from the method itself.

The paper is organised as follows. Section \ref{section:data} describes the data used and Section \ref{section:algo} gives a short overview of the Bisous filament finder. In Sections \ref{section:results} and \ref{section:discussion} we present and discuss the analysis conducted on the Bisous output.


    \begin{table}[]
        \centering
        \caption{Number of galaxies (\(N_{\rm gal}\)) and galaxy number densities (\(\rho_{\rm gal}\)) for the subsamples used in this work.}
        \begin{tabular}{l c r l}
            \hline \hline
            Subsample & mag limit & \(N_{\rm gal}\) & \(\rho_{\rm gal}\) \\
            & & & \Big(\si{Mpc\tothe{-3}}\Big) \\ \hline
            mag1 \& rnd1 & $-18.5$ & 461551 & 0.0295 \\ 
            mag2 \& rnd2 & $-19.0$ & 344509 & 0.0220 \\
            mag3 \& rnd3 & $-19.5$ & 253315 & 0.0162 \\
            mag4 \& rnd4 & $-20.0$ & 181411 & 0.0116 \\
            mag5 \& rnd5 & $-20.5$ & 114127 & 0.0073 \\
            mag6 \& rnd6 & $-21.0$ & 52215 & 0.0033 \\
            \hline
        \end{tabular}
        \label{tab:num_densities}
        \tablefoot{Subsamples named mag1-6 are obtained by using the magnitude limit in the second column. Subsamples named rnd1-6 have the same number of galaxies as in mag1-6 counterpart (as shown in the \(N_{\rm gal}\)), but the galaxies are chosen randomly from the full sample without applying the magnitude limit (see Sect \ref{section:data:sim} for more details). The simulation region in this study is a \SI[parse-numbers=false]{250^3}{Mpc^{-3}} box.}
    \end{table}

    \begin{figure}
        \centering
        \includegraphics[width=\hsize]{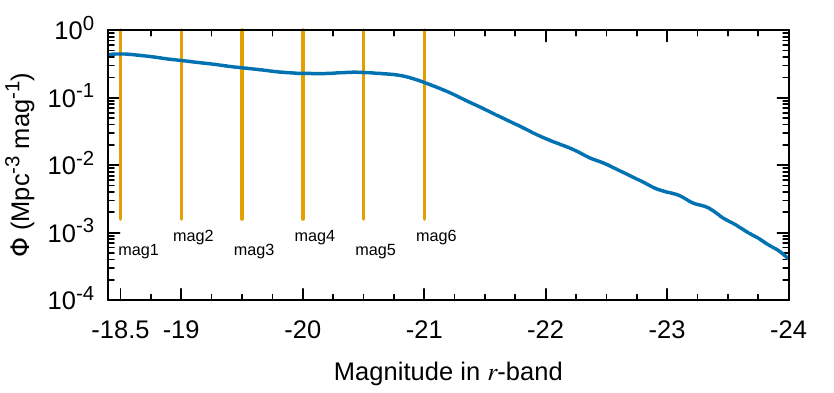}
        \caption{Luminosity function \(\Phi\) of galaxies in the \textsc{MultiDark-Galaxies} catalogue (see Subsection~\ref{section:data:sim}) with vertical lines showing the luminosity lower limit values  (from -21 to -18.5) for the subsamples used in this work (see Table~\ref{tab:num_densities}).}
        \label{fig:lum_function}
    \end{figure}
    
    \begin{figure}
        \centering
        \includegraphics[width=\hsize]{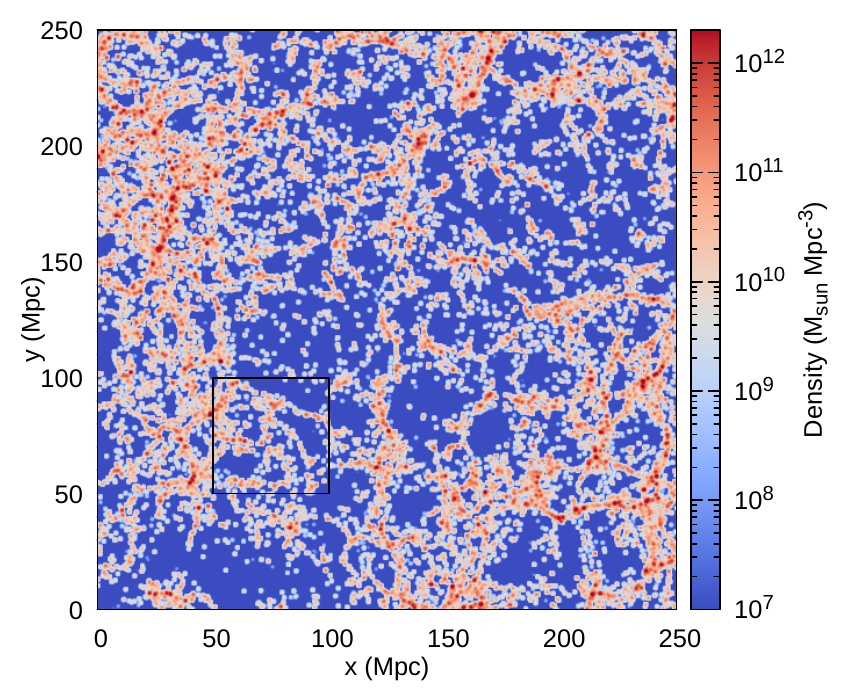}
        \includegraphics[width=\hsize]{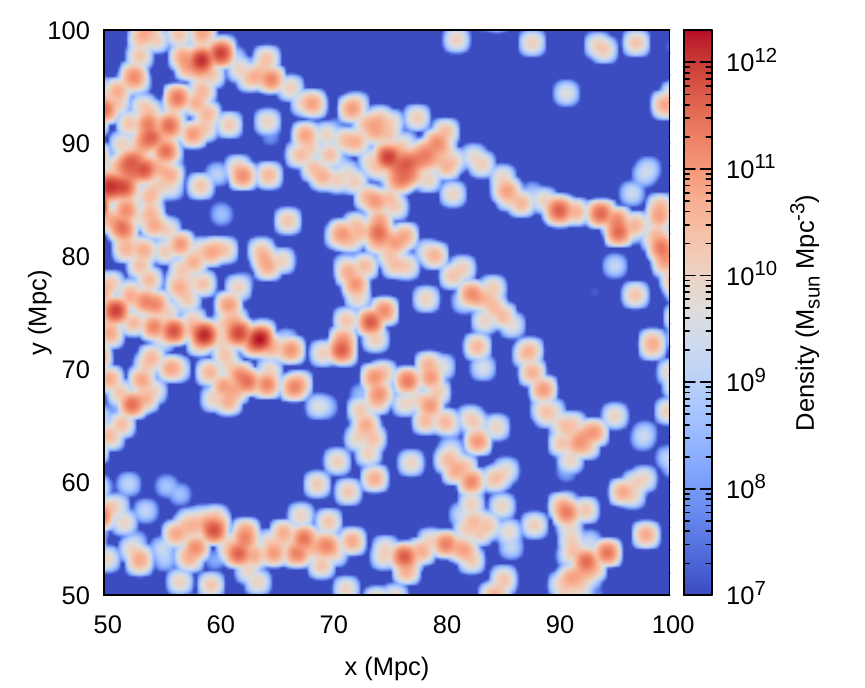}
        \caption{Mass density field of mag1 galaxies (see Table \ref{tab:num_densities}). \textit{Upper panel}: A slice of the whole data cube. The depth of the slice is \SI{10}{Mpc} and the data are smoothed with a Gaussian kernel with a width of \SI{0.7}{Mpc}. \textit{Lower panel}: Magnification of the area indicated with the square on the upper panel. For clarity, further spatial visualisations only use the smaller \SI{50 x 50}{Mpc} area, as shown on the lower panel.}
        \label{fig:mass_density_field}
    \end{figure}

\section{Data} \label{section:data}

\subsection{Simulation and galaxy catalogue}
\label{section:data:sim}

The analysis in this paper is based on a simulated mock data. For the mock data set, we used the galaxy catalogue \textsc{MultiDark-Galaxies} based on \textsc{MultiDark-Planck 2} (MDPL2\footnote{\url{https://www.cosmosim.org/cms/simulations/mdpl2/}}, \citealt{Klypin16}) simulation with \textsc{Sag} semi-analytic model for galaxies described in \cite{Knebe18}. The MDPL2 simulation is based on a dark matter-only flat \(\Lambda\) cold dark matter (\(\Lambda\)CDM) model with \textsc{Planck} cosmological parameters: \(\Omega_\mathrm{m} = 0.307, \ \Omega_\mathrm{B} = 0.048, \ \Omega_\Lambda = 0.693, \ \sigma_8 = 0.823, \ n_s = 0.96\), and \(h = 0.678\) \citep{Planck15}. The box size is \SI{1000}{\mathit{h}^{-1}\,Mpc} (\SI{1475.6}{Mpc}) with \(3840^3\) particles and mass resolution of \(m_\mathrm{p} = \SI{1.51e9}{\mathit{h}^{-1}\,\mathit{M_\odot}}\) per dark matter particle. This work only uses a smaller box of the whole simulation with a side of \SI{250}{Mpc} to have a big enough sample size for statistical analysis, but small enough volume to limit the calculation time of the Bisous filament finder (see Section~\ref{section:algo}) applied to the data.

We used \textsc{Sag} model to describe the galaxy population inside the simulation box and to obtain the input galaxy data for the Bisous model. Galaxies are embedded into identified dark matter haloes, which, in combination with the merger trees, are used to give galaxies physical properties. This semi-analytical model takes into account processes such as radiative cooling of the gas, star formation, metallicity, supernova feedback and winds, disc instability, starburst, AGN feedback, and more. Details about the applied \textsc{Sag} model can be found in \citet{Knebe18}. In our analysis, we used galaxies from the catalogue with magnitude \(\leq \! -18.5\) in the SDSS \(r\) band. We made 12 subsets from the \textsc{MultiDark-Galaxies} galaxy catalogue with 6 different galaxy number densities (see Table~\ref{tab:num_densities}). Six with magnitude cut, lower limit varying from \(-21\) to \(-18.5\)~mag (shown in Fig.~\ref{fig:lum_function} with the catalogue luminosity function for reference), and six with the same number of galaxies in the box, but the galaxies were randomly chosen from the full catalogue, instead of chosen based on the magnitude. Table \ref{tab:num_densities} gives the number of galaxies and the galaxy number density in each subsample.

Figure~\ref{fig:mass_density_field} shows the mass density field of the simulation in a \SI{10}{Mpc} thick slice. It is visible that the mass is concentrated in groups and larger structures, which form elongated filaments. The cosmic web is visible by eye, but the density field is complex and it is difficult to pinpoint the exact locations of filaments because  different methods give somewhat different results \citep{Libeskind18}.

\subsection{Observational data}

\begin{figure}
        \centering
        \includegraphics[width=\hsize]{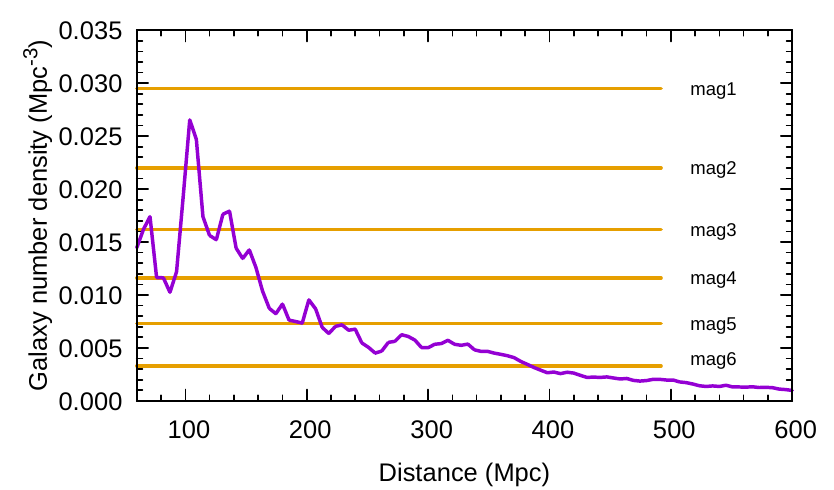}
        \caption{SDSS galaxy number density as a function of distance from the observer. The horizontal orange lines show six different galaxy number density cuts used in this work ranging from \SI{0.0033}{Mpc^{-3}} to \SI{0.0295}{Mpc^{-3}} (see Table \ref{tab:num_densities}).}
        \label{fig:gal_num_density}
    \end{figure}

For comparison with observational data, we used the SDSS DR12 \citep{SDSS_III, SDSS_DR12}. The SDSS is a flux-limited survey, meaning that the galaxy number density decreases with distance as seen from Figure~\ref{fig:gal_num_density}. We used the SDSS DR12 groups catalogue compiled by \cite{Tempel17}, where the distances are given in comoving frame and the Fingers of God effect has been removed. We chose subsamples from the mock data to emulate the galaxy number density at different distances, which are denoted in Figure~\ref{fig:gal_num_density}. The highest density sample (mag1) has a higher number density of objects than available in the SDSS. The lowest density sample (mag6) has a galaxy number density that is comparable with the galaxy number density at the distance of \SI{400}{Mpc} in the SDSS. \citet{Tempel14} shows that filaments detected from the SDSS data with the Bisous model are reliable up to the distance around \SI{200}{Mpc}, which corresponds to the mag5 subsample in this paper.

        \begin{figure}
                \centering
                \includegraphics[width=\hsize]{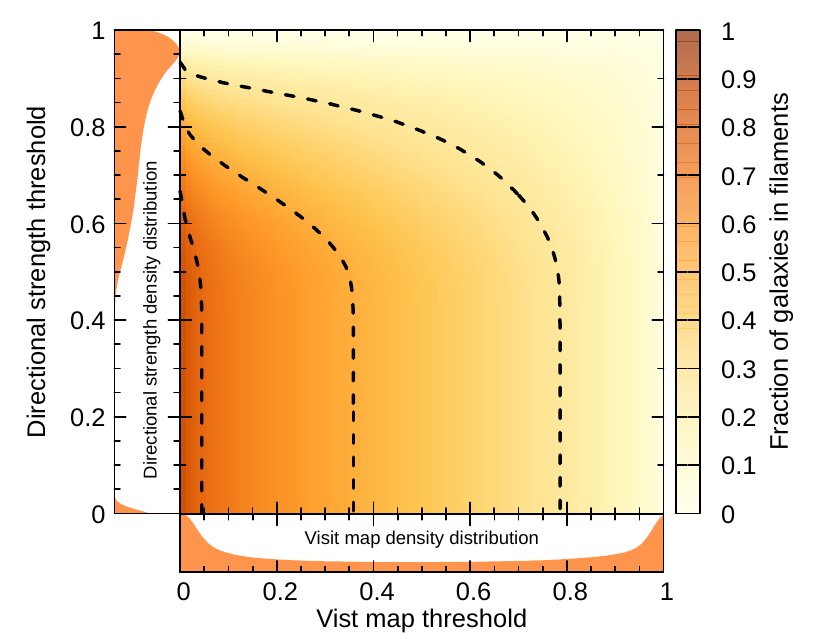}
                \caption{Fraction of galaxies in filaments with different Bisous parameters in subsample mag2. Dashed lines indicate isolines for the fraction of galaxies in filaments with values of (from left) 0.75, 0.5, 0.25. The plots on the left and at the bottom show the distributions of data with corresponding values for directional strength and visit map accordingly.}
                \label{fig:heatmap_gal}
        \end{figure}

    \begin{figure*}
        \centering
        \includegraphics[width=\hsize]{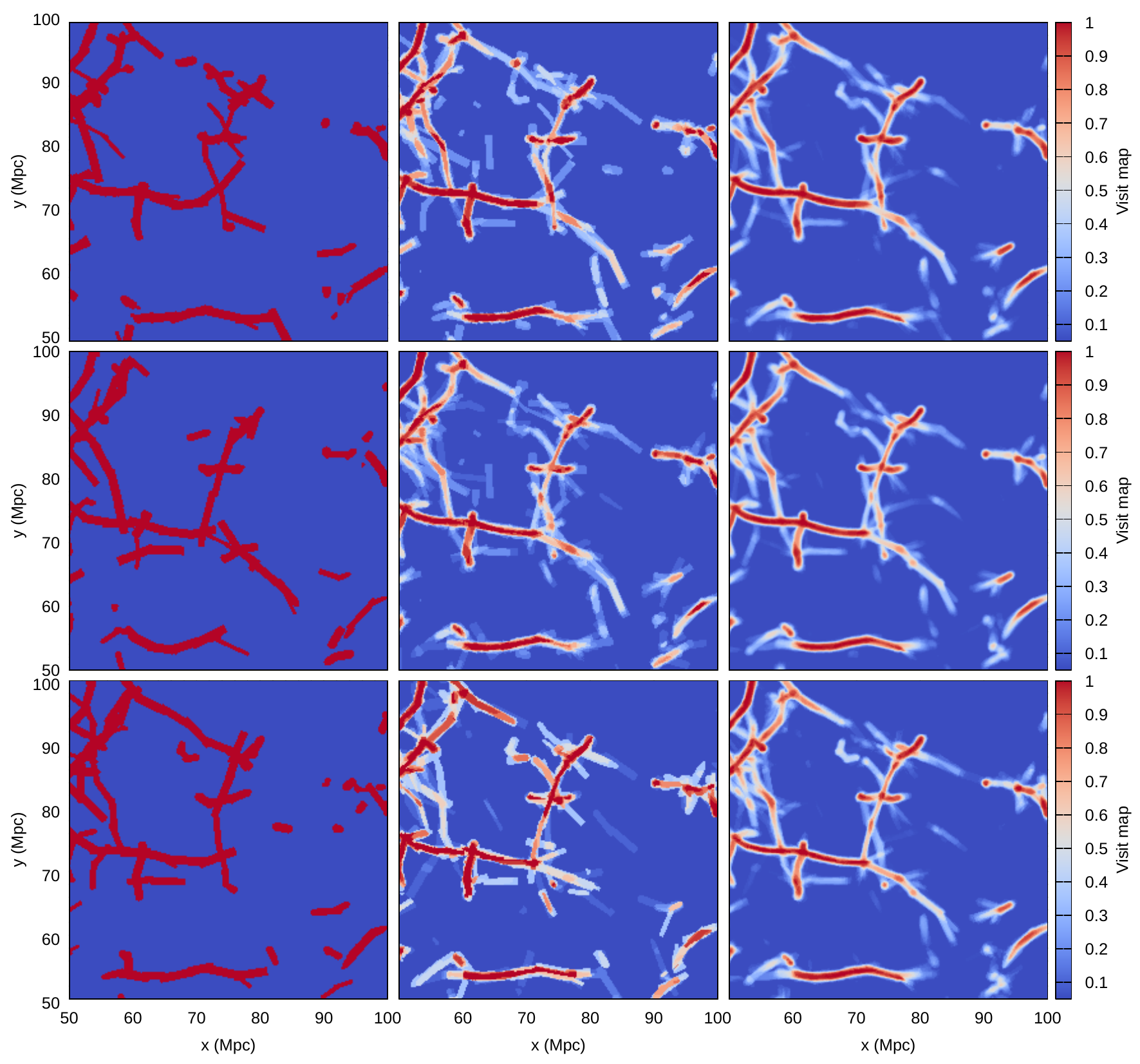}
        \caption{Maximum visit map value in a thin slice. The left column has a single realisation, the middle column 20 stacked realisations, and the right column 200 realisations of Bisous on mag2 subsample. Each row uses different realisations, which are used to construct the visit map. The thickness of slice is \SI{10}{Mpc} and for clarity, only \SI{50 x 50}{Mpc} region is shown. A single realisation can only have values of 1 or 0, and each single realisation contains only some features of the whole filamentary network. The more realisations used for the visit map the less variety there is between different visit maps. The rightmost panels have all the same features and differ only in minor details.} 
        \label{fig:single_realisation}
    \end{figure*}

\section{Filament detection algorithm}
\label{section:algo}

We used the Bisous filament finder to detect the filaments from the mock data. This finder is a stochastic tool to identify the spines of the filaments using the spatial distribution of galaxies or haloes \citep{Tempel14, Tempel16}. The Bisous has already been applied to a variety of data and has been proven to give similar results to other filament finders \citep{Libeskind18}. We give a short overview of the method below.

First, the Bisous randomly populates the volume with points with parameters (called marked points), where each point represents the centre of a cylinder and the parameters give the size and orientation of the cylinder. Each configuration of cylinders in the volume has defined energy, which depends on the position of the cylinders in relation to the underlying data of haloes and also on the interconnectedness of the filamentary network made up of the cylinders. Using the Metropolis-Hastings algorithm and the simulated annealing procedure, the Bisous model minimises the energy of the system by randomly adding, removing, or changing the cylinders.

The data of the cylinder configurations are collected over thousands of cycles, which is the basis for visit map and directional strength calculations. In general, one realisation of cylinders in the volume represents the detected filamentary network. Since the model is stochastic, the filamentary network changes from realisation to realisation. The combination of many realisations allows us to define the visit map and directional strengths that describe the detected filamentary network (see \citet{Tempel14, Tempel16} for more details).

Each coordinate has defined visit map and directional strength map value. In short, the visit map contains information on how often a coordinate in space was `visited' by a cylinder, meaning how probable it is that a random realisation has a cylinder at that position. The directional strength describes the alignment of different configurations at that position. The better the alignment, the higher the directional strength value. Both parameters range from 0~to~1.

In this work, the aforementioned maps are calculated for two different sets of points: first, in a grid with a constant step size of \SI{0.2}{Mpc} and second, for the galaxy positions in the studied volume. The grid coordinates are used to calculate the fraction of volume filled with filaments and the galaxy position coordinates emulate observations and enable us to calculate statistics for galaxies.

The filamentary structure depends on the model parameters that are used to classify areas to be in the filamentary network or not. The finder extracts filaments for a given set of points based on the set visit map and directional strength threshold values. Figure~\ref{fig:heatmap_gal} shows the fraction of galaxies inside filaments for different parameter values. As the thresholds increase the volume fill of filaments and the fraction of galaxies in filaments smoothly decrease, which depends more on the visit map threshold than on the directional strength. Figure~\ref{fig:heatmap_gal} shows an area in which the fraction of galaxies in filaments is independent of the direction strength threshold. This is because there are no points with these intermediate directional strength values and increasing or decreasing the directional strength threshold does not exclude or include any points. For the filament volume fill fraction, the independence from directional strength is much more significant because increasing the directional strength threshold excludes more area around clusters, but the volume of clusters is small compared to the total volume and therefore has little effect on volume-based statistics. The commonly adopted value for the visit map threshold is \(0.05\), which is close to the signal-to-noise ratio for a Poisson process, and for the directional strength \(0.8\). These values have been previously calibrated for the results to minimise noise, visually resemble a filamentary network, have an appropriate volume fill, and have already been used in previous works such as \citet{Tempel14}, \citet{Libeskind18}, and \citet{Kooistra2019}. This study uses the same values when extracting filaments unless explicitly stated otherwise. The visit map and directional strength map values were calculated for both the grid points and at the galaxy positions.

The full visit map and directional strength map, from which the filaments are extracted, are constructed from thousands of realisations of Bisous on the same input data. One realisation represents a configuration of the marked points (i.e. cylinders) populating the volume. Each realisation maps some, but not all the features of the filamentary network. To construct a full network, many realisations have to be stacked. The number of stacked realisations is chosen based on the density of input data and available computational resources. For example, Figure~\ref{fig:single_realisation} illustrates how stacking improves the mapping of the filamentary network. The leftmost column has only one realisation, middle column 20, and the rightmost column 200 stacked realisations; different rows use a different set of realisations for stacking. The single realisations (left column) are more coarse and incomplete in comparison to the full visit map (e.g. Fig.~\ref{fig:fil_multi}).
The rightmost column shows less difference between different compilations (different rows) than 1 or 20 realisation stacks. Also, the 200 realisation stacks already show all the features as in the full visit map and differ only in minor details.
This shows how the filamentary network is constructed from a sufficiently large number of Bisous realisations, which is usually from several hundred to a thousand realisations. Even though every single realisation has incomplete information of the full visit map, the sum of realisations produces robust results with little variance.
    

\section{Results} \label{section:results}

    \begin{figure}
        \centering
        \includegraphics[width=\hsize]{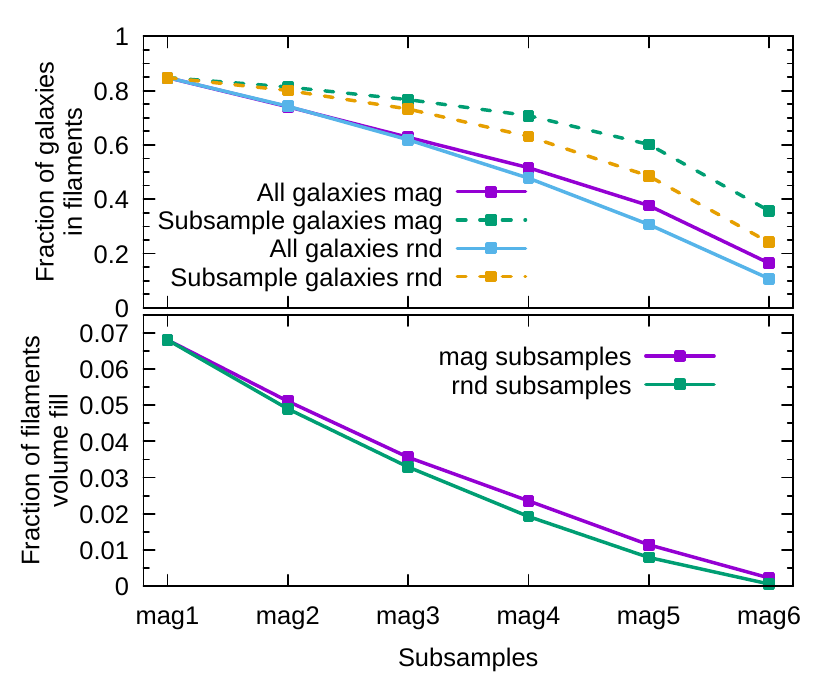}
        \caption{Fraction of galaxies inside filaments and fraction of filaments volume fill for subsets. \textit{Upper panel}: Fraction of galaxies inside filaments for different subsamples (see Table~\ref{tab:num_densities}). The solid lines show the results based on using all the galaxies in the full sample and the dashed lines using only galaxies of the subsample, that is the galaxies used to find the filaments for that subsample. \textit{Lower panel}: Filament volume filling fraction for different subsamples. Both plots show the values for the samples where cuts were done by magnitudes (mag) and also for the samples where cuts consist of randomly chosen galaxies (rnd), while keeping the same galaxy number density as the magnitude counterpart (see Table~\ref{tab:num_densities}).}
        \label{fig:fill_fractions}
    \end{figure}

Next, we present how different input data and model parameters affect the cosmic web detected by the Bisous filament finder. Subsections \ref{section:results:subsamples_intro} and \ref{section:results:fill_overlap} study the differences from inputs with different galaxy number densities and Subsection \ref{section:results:stat_err} presents the analysis on the robustness of Bisous filaments.

\subsection{Filaments from different subsamples} \label{section:results:subsamples_intro}

    \begin{figure*}
        \centering
        \includegraphics[width=\hsize]{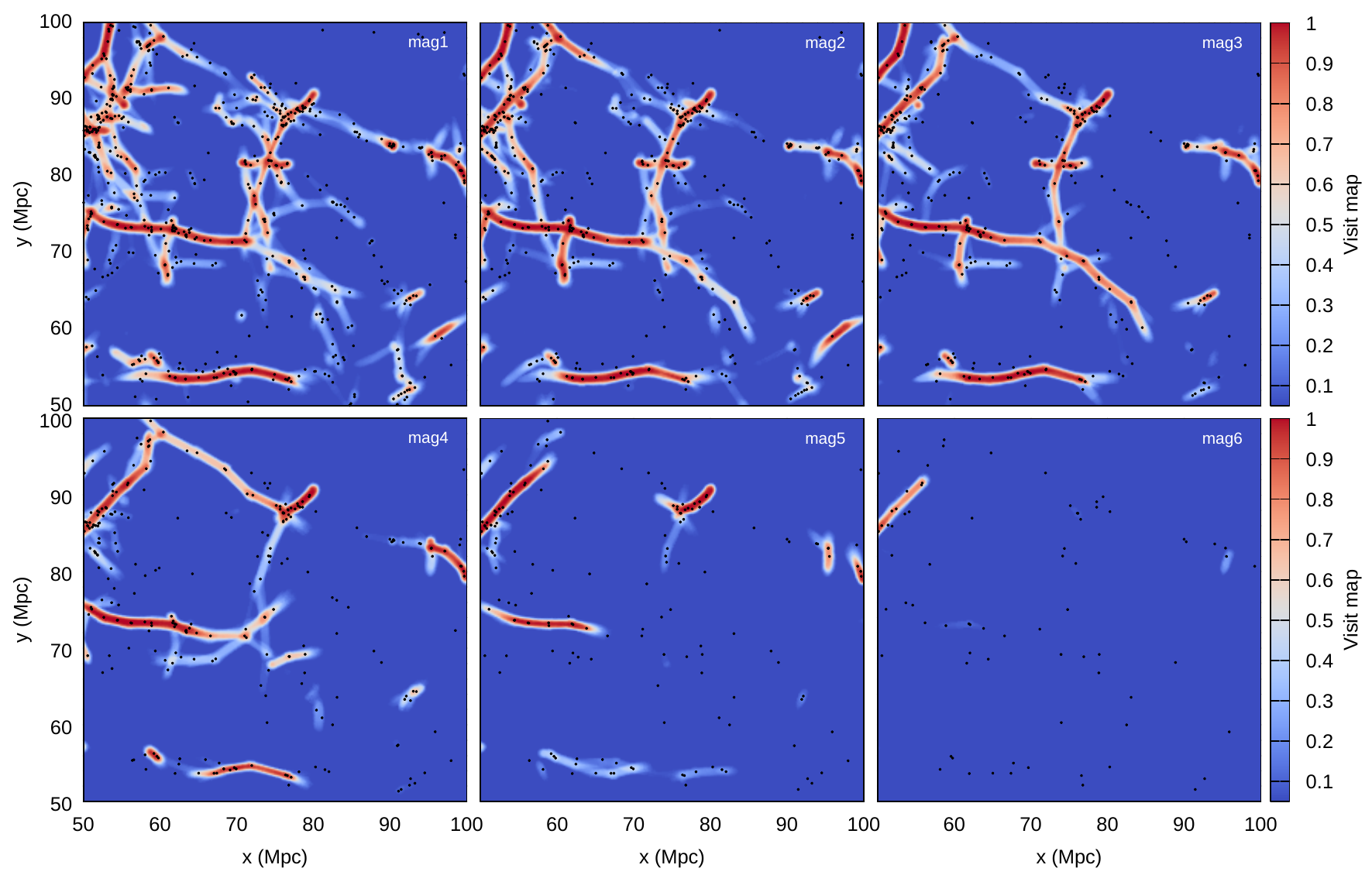}
        \caption{Visit map values at grid positions and galaxies for different subsamples. Each figure is a slice with the thickness of \SI{10}{Mpc} and the visit map value shows the maximum value over \(z\)-axis. The galaxies that are used to calculate the visit maps are shown as black dots. All the figures depict the same coordinates in the various subsamples, showing how the filamentary networks obtained from different subsamples differ visually.}
        \label{fig:fil_multi}
    \end{figure*}

Subsamples with different galaxy number densities trace the same large-scale structure, but might visually reveal slightly different cosmic web structures. In general, the lower the number density of galaxies in the subsample the lower the detected filament density (as shown in Figure~\ref{fig:fill_fractions}); this is also evident when looking directly at the visit maps (e.g. Fig.~\ref{fig:fil_multi}). This also affects how well the filaments trace the underlying dark matter density field.  With a higher number density, more filaments are detected, which better resembles the dark matter density field.

As the Bisous depends on the positions and number of galaxies, the lower-density subsamples might produce some filaments in places in which there was nothing in the higher-density subsamples. For visual reference and comparison, in Fig.~\ref{fig:fil_multi} we show all of our galaxy subsamples with their underlying visit map that the Bisous derived from those samples. The aforementioned effect of filaments from lower-density samples not appearing in higher-density samples can be seen most clearly from mag1 and mag2 (e.g. at coordinates $(x,y) = (55,55) \, \si{Mpc}$). When the galaxy number density decreases, fewer filaments are detected. The only strong filament visible in mag6 panel might be a supercluster (at coordinates \((x,y) = (55,90)\, \si{Mpc}\)), which has lots of galaxies in mag1. But these galaxies are not well aligned, therefore produce several scrambled weaker filaments; however when only keeping the brightest galaxies (as in mag6), they are concentrated on a filament, which is clearly visible.
Some areas (e.g. in mag3 around coordinates \((x,y) = (85,75)\, \si{Mpc}\)) show aligned galaxies, but no detected filament. This is because the galaxies only appear to be aligned owing to the projection effect, but are not aligned in the \(z\)-axis.

\subsection{Filaments volume fill and overlap} \label{section:results:fill_overlap}

One of the simplest ways to characterise a filament finding method is to look at how much of the volume is filled with filaments and in addition to that, if possible, how many galaxies reside in filaments. As already shown, the filament volume filling fraction decreases monotonically with decreasing galaxy number density in the sample. Analogously, the fraction of galaxies in filaments decreases monotonically as well. The underlying dark matter distribution is the same for all the subsamples and they only differ by the number of galaxies visible for the method. This only shows that the method is able to find fewer filaments from samples with low galaxy number density and does not conclude that there are fewer filaments in the volume.

In Figure~\ref{fig:fill_fractions}, the upper panel shows the fraction of galaxies in filaments. We used the same mock data set for different subsamples with different galaxy number densities. This allows us to study how many of the galaxies not visible for the filament finder were still positioned in filaments, when using other subsamples except for mag1, which has no `invisible' galaxies (e.g. for magnitude cut subsamples the galaxies fainter than the cut threshold). The upper panel of Figure~\ref{fig:fill_fractions}  shows the fraction of galaxies in filaments for both cases: only counting the galaxies that were used to detect the filamentary network (``subsample galaxies''), and counting all the galaxies (using the galaxies in mag1). To draw a comparison with observational cases, where there will always be some galaxies too faint to detect, the fraction of subsample galaxies in filaments show the fraction of observed galaxies in filaments, and the designation ``all galaxies'' also includes those that are not detected by the survey.

The lower panel of Figure~\ref{fig:fill_fractions} depicts how much of the volume is filled with filaments. A point is considered to be in a filament if its visit map and directional strength map values are over given thresholds. We used the threshold values of 0.05 and 0.8, respectively. It is important to note that the Bisous only finds spines of filaments, which does not account for the thickness of the filament, but instead uses a given radius of about \SI{1}{Mpc}, which was found to be the most suitable for filaments traced by gas in a hydrodynamical simulation \citep{Kooistra2019}. This gives an estimate of the volume fill, which might deviate from values found with methods that search filaments on different scales and thickness \citep{Libeskind18}. Using randomly chosen galaxies instead of magnitude cuts shows similar trends, but lower values overall. As these subsets also include some randomly chosen lower luminosity galaxies and have a lower fraction of galaxies in filaments, this indicates that filaments consist of preferentially higher luminosity galaxies.

        \begin{figure}
                \centering
                \includegraphics[width=\hsize]{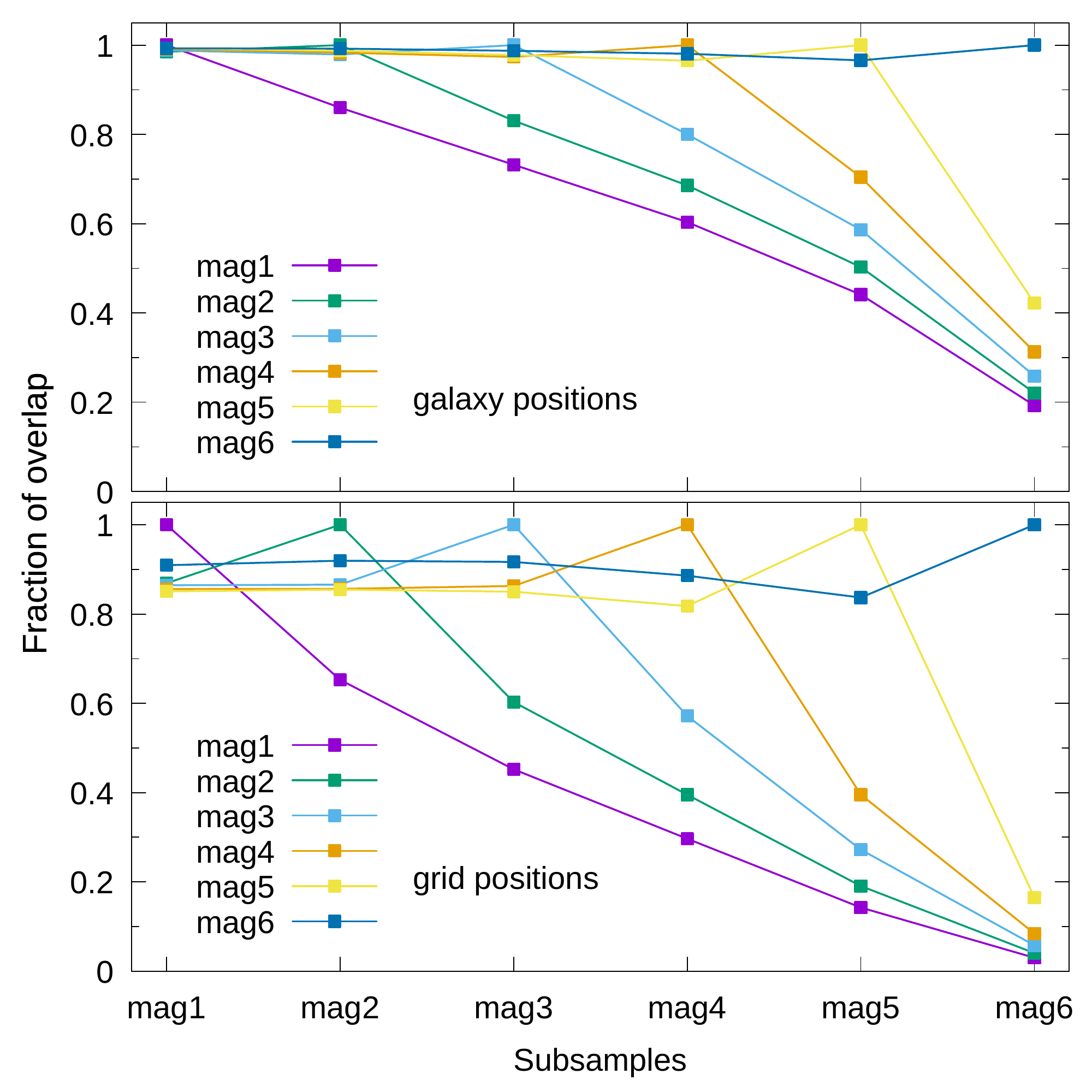}
                \caption{Overlap between filaments calculated using different subsamples compared at galaxy (\textit{upper panel}) and grid (\textit{lower panel}) positions. Lines show the fraction of subsample mag$N$ (line) filaments also present in the subsample mag$N'$ ($x$-axis). The subsamples are defined in Table~\ref{tab:num_densities}. }
                \label{fig:overlap}
        \end{figure}

        \begin{table*}[]
        \centering
        \caption{Fraction of galaxies with probability of being in a filament close to 0 or 1 and with different visit map thresholds (\(vm_{\rm{t}}\)).}
        \begin{tabular}{c c c c c c c}
            \hline \hline
            \(vm_{\rm t}\) & \multicolumn{6}{c}{Probability threshold}\\
            & 0.001 & 0.01 & 0.02 & 0.05 & 0.1 & 0.317 \\ \hline
            0.03 & \(_{0.19} \, 0.38 \, _{0.19}\) & \(_{0.23} \, 0.46 \, _{0.23}\) & \(_{0.26} \, 0.51 \, _{0.25}\) & \(_{0.30} \, 0.58 \, _{0.28}\) & \(_{0.35} \, 0.67 \, _{0.32}\) & \(_{0.48} \, 0.88 \, _{0.40}\) \\ 
            0.05 & \(_{0.20} \, 0.38 \, _{0.18}\) & \(_{0.25} \, 0.47 \, _{0.22}\) & \(_{0.28} \, 0.52 \, _{0.24}\) & \(_{0.33} \, 0.60 \, _{0.27}\) & \(_{0.38} \, 0.68 \, _{0.30}\) & \(_{0.51} \, 0.89 \, _{0.38}\) \\
            0.08 & \(_{0.26} \, 0.44 \, _{0.18}\) & \(_{0.30} \, 0.51 \, _{0.21}\) & \(_{0.33} \, 0.56 \, _{0.23}\) & \(_{0.37} \, 0.63 \, _{0.26}\) & \(_{0.43} \, 0.72 \, _{0.29}\) & \(_{0.54} \, 0.90 \, _{0.36}\) \\
            0.12 & \(_{0.30} \, 0.47 \, _{0.17}\) & \(_{0.36} \, 0.56 \, _{0.20}\) & \(_{0.38} \, 0.60 \, _{0.22}\) & \(_{0.43} \, 0.67 \, _{0.24}\) & \(_{0.48} \, 0.75 \, _{0.27}\) & \(_{0.57} \, 0.91 \, _{0.34}\) \\
            \hline
        \end{tabular}
        \label{tab:tail_fractions}
        \tablefoot{Directional strength threshold was \(ds_{\rm t} = 0.8\) for every run, the visit map threshold for each is shown in the \(vm_{\rm t}\) column, and the commonly used value is \(vm_{\rm t} = 0.05\). The probability threshold shows the maximum difference, which is considered "close" to 0 or 1. These sizes correspond to a galaxy being classified the same in 99.9\%, 99\%, 98\%, 95\%, 90\%, and 68.3\% of runs. Each cell in the table first shows the fraction of galaxies with probability close to 0 (in small font), next the sum of the fractions with probabilities close to both 0 and 1, and then the fraction of galaxies with probability close to 1 (in small font). Majority of the galaxies have either a very low or very high probability being inside a filament.}
    \end{table*}

For a more thorough comparison between different subsets, we studied how the networks of filaments differ from subset to subset, that is how many of the coordinates (either in grid or at galaxies positions) have the same classification between different subsets. Figure~\ref{fig:overlap} shows the overlap of coordinates classified as filaments in different subsamples. The upper panel compares the galaxy position coordinates. As shown in the figure, more than 97\% of the galaxies that are classified to be inside filaments in the lower-density subsamples are also inside filaments in higher-density subsamples. This means that if the filament finder classifies a galaxy to be inside a filament even with incomplete data, then there is a very high probability that the galaxy is also in a filament with more complete data.

The fixed grid position analysis (Fig.~\ref{fig:overlap} lower panel) shows that in addition to lower-density subsamples having smaller filament fill fractions, there are also some new areas classified as filaments, that is new random filaments that are not present in higher-density samples (as mentioned in Sect.~\ref{section:results:subsamples_intro}). About 10-15\% of the points in lower-density subsamples classified as filaments are not filaments in the highest-density sample.

There is no ground truth as to what the definition of a filament is nor is there a method to detect the true filaments. But when combining the overlapping of filaments in different subsamples (Fig.~\ref{fig:overlap}) with the visual representation of filaments (Fig.~\ref{fig:fil_multi}), we can conclude that when moving from higher- to lower-density subsamples, the probability to detect the filamentary network decreases starting with less populated regions;  only the most prominent features of the network can be detected. In contrast, if we start with the low-density subsamples, then the filaments detected with those galaxies are mostly present in filamentary networks detected with higher-density subsamples. About 85\% of the detected filaments are reliable. As the fractions of reliable filaments stay stable for all the subsamples we used, we expect this trend to persist with even higher galaxy number densities.

\subsection{Analysis of statistical reliability of the Bisous model} \label{section:results:stat_err}

As already mentioned, the Bisous model is stochastic and therefore does not give deterministic results for a given input. We conducted a statistical analysis on the subsample mag2, which is comparable with galaxy number density in SDSS at \SI{100}{Mpc}. A full Bisous model run on a single core with 350 000 galaxies takes a few days. Each core can run a separate Bisous process and thus have a very simple parallelisation.

For this subsample, we calculated the underlying large-scale structure \(N_R=200\) times, giving us 200 full Bisous runs \(R\), meaning each galaxy has \(N_R\) different visit map values. We defined the probability of being in a filament for a galaxy as
\begin{equation}
    P = \frac{\sum\limits_{R=1}^{N_R} \mathbb{1} \{ vm_R \geq vm_{\rm t} \wedge ds_R \geq ds_{\rm t} \}}{N_R} \ ,
\end{equation}
where the numerator is a sum over all the Bisous runs and \(\mathbb{1} \{ vm_R \geq vm_{\rm t} \wedge ds_R \geq ds_{\rm t} \}\) is an indicator function, which returns 1 for every run \(R\), where \(vm_R\) and \(ds_R\) are higher than the chosen thresholds \(vm_{\rm t}\) and \(ds_{\rm t}\). A visit map threshold of \(vm_{\rm t} = 0.05\) and a directional strength threshold of \(ds_{\rm t} = 0.8\) are used unless explicitly stated otherwise. Table~\ref{tab:tail_fractions} shows the fraction of galaxies that have the probability of being in a filament close to 0 or 1, that is they are classified the same in almost all the runs. Chosen probability thresholds correspond to a galaxy being classified the same in 99.9\%, 99\%, 98\%, 95\%, 90\%, and 68.3\% of runs. With 95\% reliability, about 60\% of the galaxies have the same classification every time the Bisous model is run on the same input data. This value depends on the number of realisations stacked and more realisations means a higher confidence of the classification. Table~\ref{tab:tail_fractions} shows the higher the visit map threshold value the less uncertain galaxies and the more galaxies not in filaments. Increasing the visit map threshold reduces the noise (uncertain galaxies)s but also changes values of galaxies that should be in a filament, which is undesirable. Table~\ref{tab:tail_fractions} gives estimates on the reliability of the Bisous model to classify a single galaxy and this does not have a one-to-one correspondence with the reliability of the method itself.

\begin{figure}
                \centering
                \includegraphics[width=\hsize]{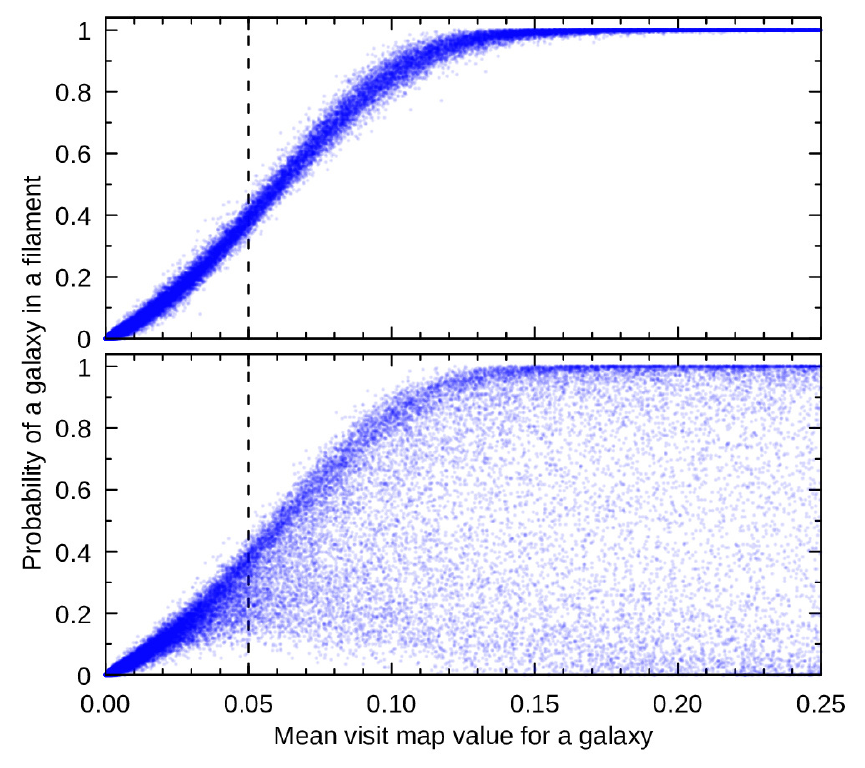}
                \caption{Probability of being in filament as a function of the mean visit map value of a galaxy. The dashed lines show the visit map threshold value. \textit{Upper panel}: Probability of being in a filament is calculated using only the visit map value. \textit{Lower panel}: Probability also accounts for the directional strength threshold. The outliers come from clusters, in which  visit map values are high, but directional strength is weak because the filament orientation is not clearly defined inside clusters.}
                \label{fig:fil_prob}
        \end{figure}

Figure~\ref{fig:fil_prob} illustrates the relationship between the probability of being in a filament and the mean visit map value of a galaxy over all the $N_R$ runs. In the upper panel, where only the visit map is used for defining filaments (i.e. \(ds_{\rm t} = 0\)), the relation is almost linear until the probability achieves a value 1. This means that if a galaxy has a visit map value that is higher than approximately 0.14, then the galaxy is in a filament almost every time the Bisous model is run on this input data. And when a galaxy has a visit map value of 0.05, which is the edge case for being considered to be in a filament, then there is \SI{\sim 40}{\percent} chance that the galaxy is also in a filament any other time the Bisous is run on the same data. The lower panel in the same figure has a different definition for the probability of being in a filament because the definition also accounts for the directional strength threshold of \(ds_{\rm t} = 0.8\). The diverging data points are galaxies inside clusters. Clusters have high visit map values, but the orientation of filaments is hard to define and therefore clusters have low directional strength values. As seen from the figure, low visit map values have almost no outliers because clusters generally have high visit map values; the area around the mean visit map value of 0.05 and the probability of 0 is empty, but higher visit map values have galaxies with no probability of being in a filament. The majority of galaxies still follow the same function defined by the probability without accounting for directional strength; as shown in Table~\ref{tab:tail_fractions}, most of the galaxies have a probability close to either 0 or 1.

\begin{figure}
                \centering
                \includegraphics[width=\hsize]{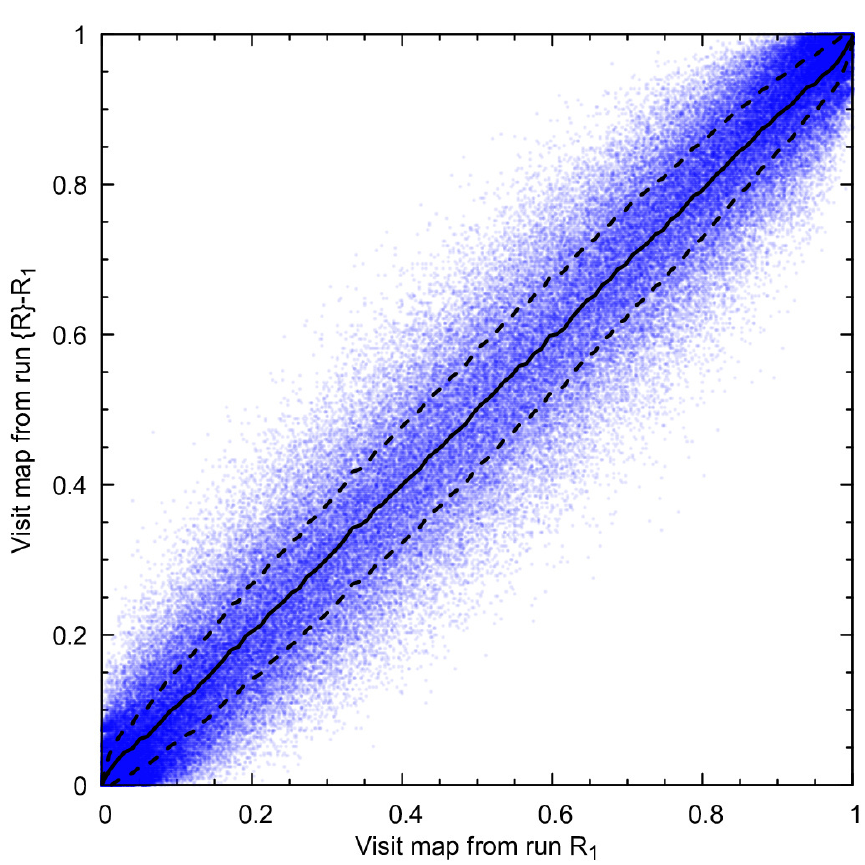}
                \caption{Comparison of the visit map values of a galaxy in different Bisous runs \(R\). The horizontal axis has the visit map value of the galaxy from the first run \(R_1\), and all the other runs \(\{R_2 \dots R_{N_R}\}\) are scattered using their visit map value for the vertical axis and the first run for the horizontal axis. Every point on the figure represents a visit map value from a single run for a single galaxy. The solid line represents the mean and dashed lines \(\pm \sigma\) values. Both mean and standard deviations are smoothed for visual clarity. The Pearson correlation coefficient is \(R=0.98\).}
                \label{fig:vmap_comp}
        \end{figure}

To characterise the reliability of subsequent runs with the same data and parameters, we studied the correlation of 200 different runs. In Figure~\ref{fig:vmap_comp} all the galaxies are ordered by the visit map value of the first run, and all the other visit map values for the galaxy are shown. As the run for ordering (\(x\)-axis) is chosen arbitrarily (first run for every galaxy), which could be the outlier data point for that galaxy, then the mean and standard deviation values had to be smoothed to suppress noise. The Pearson correlation coefficient for the statistical analysis is \(R=0.98\), showing a very strong correlation between different runs on the same data. This shows little variance in the results obtained from different runs with the same inputs and indicates good reliability between the runs for the Bisous model.

\section{Discussion and conclusions}
\label{section:discussion}

The main feature of the Bisous filament finder is the ability to detect the cosmic web by only requiring the galaxy positions as input data. This means it is easy to apply to observational data from surveys, as there is no need to reconstruct fields (e.g. velocity or tidal fields) or acquire additional information besides positions. Still, the Bisous model has been shown to trace, for example, the luminosity density field \citep{Tempel14} and the underlying dark matter velocity field \citep{Libeskind2015}. This shows that the model is able to trace the large-scale structure, which is shaped by the underlying dark matter distribution, using the distribution of galaxies.

This work studies the reliability of the Bisous filament finder by applying it to different cuts from a simulation to study how the method works with input data that have different galaxy number densities, but are otherwise the same, as well as how the results change when the parameters of the model are changed. For the input data, we used galaxies from the dark matter-only \(\Lambda\)CDM model simulation \textsc{MultiDark-Planck 2} with the \textsc{Sag} semi-analytical model for galaxies. A set of 12 subsamples with six different galaxy number densities was constructed from the simulation galaxy catalogue (see Table~\ref{tab:num_densities}). One set of six subsamples was obtained via magnitude cuts and the other set of six subsamples was chosen to have the same galaxy number densities as the counterpart magnitude cut subsamples, but galaxies were appointed by random choice rather than by magnitude. These randomly constructed subsamples were used to study any magnitude dependent effect that might not be apparent in the set of subsamples with a magnitude cut. For the results to be comparable to studies with observational data, the galaxy number densities in different subsamples approximately correspond to the SDSS densities for distances from \SI{100}{Mpc} up to \SI{400}{Mpc} (see Fig.~\ref{fig:gal_num_density}).

The Bisous model uses a marked point process with interactions to trace the filamentary network. This tool uses galaxy positions and stochastic methods to assess how well a proposed filamentary network fits the underlying galaxy distribution and also the connectivity of the network itself. Based on this assessment the method tries to optimise the proposed network. Averaging over a large number of the Bisous realisations on the same data reveals the filamentary network traced by the input data.

Varying the visit map, directional strength map thresholds, and the parameters of the model shows that the volume fill and the fraction of galaxies in filaments depend more on the visit map value (see Fig.~\ref{fig:heatmap_gal}). Often, the visit map value is already a good enough tracer for the filamentary network. The main advantage to using the directional strength parameter is to exclude clusters from the filamentary network because galaxies in clusters have a high visit map value, but the filaments have no preferential direction inside clusters and so the directional strength values are low. Increasing the threshold value for directional strength removes a larger volume around the cluster from the filamentary network. As clusters account for little to total volume, the directional strength has little affect on the fraction of volume filled with filaments, but the effects are more noticeable for the fraction of galaxies in filaments.

Comparing subsamples with different galaxy number densities shows a smooth decline in the fraction of volume filled by filaments and the fraction of galaxies in filaments (see Fig.~\ref{fig:fill_fractions}). The former ranges from 7\% to almost 0\% and the latter from 80\% to 15\%. The fraction of filament volume fill is the simplest output parameter by which to compare different filament finders to each other (see comparison of different filament finders in \citealt{Libeskind18}). When the galaxies in the subsamples are chosen randomly, the fraction of galaxies is consistently lower than in the magnitude cut subsamples. This shows that the higher luminosity galaxies are preferentially located in filaments.

We analysed the persistence of the features of the cosmic web found with the Bisous model. For this, we compared features found with lower galaxy number density input data to those with higher galaxy number density input data and quantified how many of the features were still present and how many were changed by the new data.
We found that galaxies that were in filaments in lower-density subsamples were \num{>97}\% of the time also in filaments in higher-density subsamples (see Fig.~\ref{fig:overlap} upper panel). Approximately 85\% of the filaments found by the Bisous would be reliable filaments that would persist in the results when we increased the galaxy number density (see Fig.~\ref{fig:overlap} lower panel). This indicates that the reliability of the method when applied to incomplete data (e.g. observations) is reasonably good. A lower galaxy number density in input data means that the Bisous model finds fewer filaments, but the filaments it does find are quite robust. These results give quantitative measures to assess the robustness of filaments detected with the Bisous model.

Lastly, we ran the Bisous on the same input data with the same parameters 200 times. Approximately 60\% of the galaxies are classified the same, either in a filament or not in a filament, 95\% of times the Bisous model is run. But if we consider only 68\% of the runs, more than 90\% of the galaxies are classified the same (see Table~\ref{tab:tail_fractions}). The correlation between different runs, assessed by the visit map values assigned to galaxies, is \(R=0.98\) (see Fig.~\ref{fig:vmap_comp}). As different runs show little variance, this means the statistical error originating from the stochasticity of the method is mitigated by averaging the Bisous results over a large number of realisations, and the results are robust.

These results help to illustrate the reliability of the Bisous filament finder output. The results also help to assess the required density of input data, for example the target number density in surveys, for the method to give reasonable results.

The Bisous has been applied to SDSS, the catalogue of filaments, and an overview of the process is presented in \citet{Tempel14}. We will develop the model further to combine the spectroscopic and photometric galaxies for improved performance. \citet{Kruuse2019} shows that the photometric galaxies correlate with the Bisous filaments, and as shown in this paper, the higher the input number density the higher the quality of the detected filamentary network. For example, in SDSS DR12, the photometric galaxies outnumber the spectroscopic galaxies by 100 to 1 \citep{Beck2016}. The Bisous model would greatly benefit from increasing the usable galaxy number density and therefore increasing the accuracy of the filaments detected by spectroscopic data. We aim to apply the improved Bisous model on the upcoming narrow-band photometric redshift data from Javalambre Physics of the Accelerating Universe Astrophysical Survey (J-PAS) \citep{JPAS, miniJPAS} in combination with spectroscopic data from SDSS.

\begin{acknowledgements}
We thank the referee for their detailed feedback and suggested improvements, and Jukka Nevalainen for useful comments that helped clarify and improve the contents of this paper. Part of this work was supported by institutional research funding IUT40-2, PRG1006 of the Estonian Ministry of Education and Research. We acknowledge the support by the Centre of Excellence “Dark Side of the Universe” (TK133) and by the grant MOBTP86 financed by the European Union through the European Regional Development Fund.

The CosmoSim database used in this paper is a service by the Leibniz-Institutefor  Astrophysics  Potsdam  (AIP). The \textsc{MultiDark} database was developed in cooperation with the Spanish MultiDark Consolider Project CSD2009-00064. The authors gratefully acknowledge the Gauss Centre for Supercomputing e.V. (www.gauss-centre.eu) and the Partnership for Advanced Supercomputing in Europe (PRACE, www.prace-ri.eu) for funding the \textsc{MultiDark} simulation project by providing computing time on the GCS Supercomputer SuperMUC at Leibniz Supercomputing Centre (LRZ, www.lrz.de).

For plotting, we used gnuplot-colorbrewer colour schemes available at \url{https://github.com/aschn/gnuplot-colorbrewer}, and for data analysis, we used the Julia Programming Language \citep{julialang}.
\end{acknowledgements}

\bibliographystyle{aa}
\bibliography{Muru_BIB_AA_2020_39169}

\end{document}